# Entropic measure of spatial disorder for systems of finite-sized objects

R. Piasecki [*]

*Institute of Chemistry, University of Opole, Oleska 48, PL 45052 Opole, Poland*

**Abstract**

We consider the *relative* configurational entropy per cell $S_\Delta$ as a measure of the degree of spatial disorder for systems of finite-sized objects. It is highly sensitive to deviations from the most spatially ordered reference configuration of the objects. When applied to a given binary image it provides the *quantitatively* correct results in comparison to its point object version. On examples of simple cluster configurations, two-dimensional Sierpiński carpets and population of interacting particles, the behaviour of $S_\Delta$ is compared with the normalized information entropy $H'$ introduced by Van Siclen [Phys. Rev. E 56, (1997) 5211]. For the latter example, the additional middle-scale features revealed by our measure may indicate for the traces of self-similar structure of the weakly ramified clusters. In the thermodynamic limit, the formula for $S_\Delta$ is also given.

## 1. Introduction

The problem of finding static morphological measures suitable to the quantitative characterization of complex microstructures was considered from alternative viewpoints and using more or less subtle mathematics. The family of so-called Minkowski functionals [1] defined within the integral geometry approach or much more specialized measure of disorder of labyrinthine patterns [2,3] as well as general *n*-point distribution function formalism [4] are examples of methods rather difficult in practical applications.

---

[*] *E-mail:* piaser@uni.opole.pl

Thus, much research effort has also been devoted for developing simple tools for searching correlation between the macroscopic properties and microstructure attributes of the medium. Recently, a comprehensive review devoted to that point for porous structures has been given by Hilfer [5]. Also, the similar problem of determining the effective properties of random heterogeneous media from the morphology was shortly reviewed by Torquato [6].

Here we will focus on such measure that can be easily applied to a digitized image of the microstructure or computer generated pixel distributions. Usually, after subdivision of binary image into equal square cells the analyzed objects are approximated by a distribution of point markers [7,8]. By point objects we understand objects small enough in comparison to the cell size $k \times k$ expressed in pixels. This model situation has been assumed in Refs. [9-12]. However, it has been shown [13] that this idealized for black-white micrographs method needs a modification in the case of finite-sized objects like pixels. Such a modification leads to *quantitatively* correct results. Nevertheless, from a *qualitative* point of view using of the point measure for binary images is still acceptable. It should be also stressed that there are applications suitable for point measures only.

Recently, a novel approach based on the adaptation of Shannon information entropy has been developed as the "local porosity entropy" [14] and the "configuration entropy" [15,16] concepts. These two entropic measures, worked out to characterize random microstructures represented in micrographs or digitized images, were found to be rigorously connected [17]. Then Van Siclen's interesting study [18], briefly outlined in Appendix A, has proposed the quantitative characterization of microstructure inhomogeneity by the "normalized information entropy" *H'*. Van Siclen's work has motivated us to refresh our idea of using a linear transformation of configurational entropy, mentioned in [11], as a physical measure of the degree of spatial inhomogeneity. However, only the variant for point objects has been recently reported [19].

The purpose of this paper is to provide the entropic measure $S_\Delta$ applicable for systems of equally sized objects. Binary images, where black pixels play the role of indistinguishable "particles", are the systems of interest. Our proposition, specified in Section 2, modifies the point object measure [12, 19] to the case of finite-sized objects. It differs from the other entropic approaches mentioned above. The measure is obtained for every length scale by subtracting the configurational entropy for a given arrangement of black pixels from the entropy corresponding to the most spatially ordered reference configuration and dividing the difference by the number of cells. For a given number of objects, such an approach provides a sensitive and *quantitatively* correct comparative characterization of digitized two-phase microstructures at every length scale. In the thermodynamic limit, the simple expression for this measure is also obtained.

To illustrate the basic features of $S_\Delta$ and for their comparison with $H'$, the simulated distributions beginning from the simple cluster arrangements through two-dimensional Sierpiński carpets up to the population of interacting particles (chosen from Van Siclen's paper [18]), are presented in Section 3. While for the random compact aggregates of particles and weakly ramified clusters of the interacting particles, the length scale at which the first well shaped peak of $S_\Delta$ appears corresponds to the first maximum in $H'$ indicating for clustering of the objects, the sequential $S_\Delta$ peaks of various heights for Sierpiński carpets and partially for the population of interacting particles, contain more intricate information in comparison with smoothly shaped $H'$. In the final section we make concluding remarks and indicate some open problems.

## 2. Relative configurational entropy per cell

Let a binary image of size $L \times L$ in pixels be treated as a set of $n$ indistinguishable finite-sized objects, that is black pixels of size $1 \times 1$ representing "particles" of a system and randomly distributed in $\chi$ numbered lattice cells of size $k \times k$. The image area equals to the total number $\chi_0 \equiv \chi k^2$ of the unit cells of size $1 \times 1$. For the nontrivial binary images the particle number $n$ satisfies the inequality $0 < n < \chi_0$ and hence, the particle fraction $\varphi \equiv n/\chi_0$ holds the relation $0 < \varphi < 1$. For each length scale $k \equiv (L \times L/\chi)^{1/2}$ with a given distribution $(n_1,..., n_i,..., n_\chi)$, i.e. having fixed numbers $n_i$ of particles in $i$th cell and fulfilling the two constraints, $n_1 + n_2 + ... + n_\chi = n$ and $n_i \leq k^2$, one can associate a configurational entropy $S(k) = k_B \cdot \ln \Omega(k)$, where the Boltzmann constant will be set to $k_B = 1$ for convenience. Note that the length scale $k$ is equal to the lattice cell side-length. The number $\Omega$ of distinguishable arrangements of the particles, that is the number of possible ways of generating the fixed distribution regarded as configurational macrostate, is given by

$$\Omega(k) = \prod_{i=1}^{\chi} \binom{k^2}{n_i}. \tag{1}$$

For easy reference we recall here the full formula of [13] for the probability of configurational macrostate appearance

$$P_{n_1,\ldots,n_\chi}^{n,\,k\geq 1} = \frac{\prod_{i=1}^{\chi}\binom{k^2}{n_i}}{\binom{\chi_0}{n}}. \tag{2}$$

Indeed, at scale $k > 1$ any macrostate $(n_1,\ldots, n_i,\ldots, n_\chi)$ with $n_i \leq k^2$ can be realized by a number of distinguishable spatial arrangements of $n$ particles placed into the numbered unit cells, i.e., some kind of configurational microstates $(n_1,\ldots, n_j,\ldots, n_{\chi_0})$ at scale $k = 1$ with $n_j \in \{0, 1\}$. Taking into account the local spatial arrangements in each of $\chi$ cells and assuming that every microstate is equally likely, the ratio of the number of the proper microstates to the number of all possible microstates gives us the above formula.

This formula holds also for any polygon composed of squares. For a rectangular binary image of size $L \times 2L$ and $k \times k \equiv L \times L$ we have $\chi = 2$, and the formula of Ref. [18] (cf. Eq. (1)), also given in Appendix A as (A1), is formally recovered. However, that formula was applied to a different physical situation, that is the evaluation of probability of finding exactly $i$ particles in a $k \times k$ square region if $N$ particles were randomly distributed over a system of size $L \times L$. To show that correspondence we replace $k \to L$, $n \to 2n$, $\chi_0 \to 2L^2$ in our expression (2) as well as $k \to L$, $N \to 2N$, $L^2 \to 2L^2$ in Van Siclen's formula. Obviously, it is clear that the following correspondence of the symbols $n \leftrightarrow N$ and $n_i \leftrightarrow i$ also holds.

The highest possible value of configurational entropy, $S_{max}(k) = \ln \Omega_{max}(k)$, is related to the most spatially ordered object configuration at a given length scale. We shall call such a configuration the reference configurational macrostate (RCM). No other distribution for a given binary image has a higher degree of spatial uniformity at a given length scale than the appropriate RCM. The simplest description of RCM is given by the following condition: for each pair $i \neq j$ must be $|n_i - n_j| \leq 1$. Its correctness was confirmed by numerical simulations. The same rule is satisfied for the point object approach [19]. Thus, the maximal number of the proper configurational microstates is

$$\Omega_{max}(k) = \binom{k^2}{n_0}^{\chi - r_0} \cdot \binom{k^2}{n_0 + 1}^{r_0}, \qquad (3)$$

where $r_0 = n \bmod \chi$, $r_0 \in 0, 1, ..., \chi - 1$ and $n_0 = (n - r_0)/\chi$, $n_0 \in 0, 1, ..., k^2 - 1$.

To evaluate for each length scale $k$ the deviation of the actual configuration from the appropriate RCM it is natural to consider the difference $S_{max}(k) - S(k)$. Computer simulations show that after averaging $S_{max}(k) - S(k)$ over the number of cells $\chi$, a high sensitivity to spatial particle arrangements at each length scale $k$ is revealed. Moreover, this averaging procedure is necessary to obtain the crucial property of the measure (point 6 discussed below) allowing for its calculation at every length scale. Therefore, we define the measure as $S_\Delta(k) \equiv [S_{max}(k) - S(k)]/\chi$. Its final form can be written as follows:

$$S_\Delta(k) = \frac{r_0}{\chi} \ln\left[\frac{k^2 - n_0}{n_0 + 1}\right] + \frac{1}{\chi} \sum_{i=1}^{\chi} \ln\left[\frac{n_i!\,(k^2 - n_i)!}{n_0!\,(k^2 - n_0)!}\right]. \qquad (4)$$

The measure $S_\Delta(k)$ exhibits a number of useful poperties illustrated in the next section:

(1) According to definition of the measure its lowest value equals to 0 and it is always reached at boundary length scales: $S_\Delta(k = 1) = 0$ and $S_\Delta(k = L) = 0$. Otherwise, the value $S_\Delta(k) = 0$ indicates that the configurational macrostate is equivalent to the appropriate RCM, when for each pair i ≠ j we have $|n_i - n_j| \leq 1$. For a system perfectly ordered at a given length scale we have $r_0 = 0$ and each $n_i = n_0 \equiv n/\chi$.

(2) The highest value for given length scale appears when each of $[n - (n \bmod k^2)]/k^2$ cells is fully occupied and at most one cell can be partially filled by $n \bmod k^2$ particles. Such a configurational macrostate corresponds to the strongest deviation per cell from the appropriate RCM and can be termed the maximally disordered state of the system for the length scale considered.

(3) The first well-shaped peak of $S_\Delta(k)$ corresponds to such configurations where the cells strongly occupied, $n_i \gg n_0$, as well as weakly filled, $n_i \ll n_0$, dominate. Generally, we can say that the tendency to cluster is marked in such a length scale. The repetiting decreasing maxima indicate for the clustering of clusters. The constant distance between the sequential peaks of various height is typical for Sierpiński carpets while the increasing intervals appear for patterns of grouped clusters of similar sizes.

(4) The deep minima in $S_\Delta(k)$ describe relatively more ordered configurations where the dominant contribution comes from the cells occupied by $n_i \approx n_0$ particles. The sequential well marked minima appear at length scales which reflect periodicity of the whole microstructure.

(5) The height of maxima and depth of minima allow for the comparison of relative intensity of deviations per cell from the appropriate RCM since the total number $n$ of particles is conserved for the analyzed pattern.

(6) The following property allows us to calculate the value of the measure at every length scale $k$. If the final pattern of size $mL \times mL$, where $m$ is a natural number, is formed by periodical repetition of an initial arrangement of size $L \times L$ then the value of the measure at a given length scale $k$ (commensurate with the side length $L$) is *unchanged* under the replacement $L \times L \leftrightarrow mL \times mL$ since it also causes $n \leftrightarrow m^2 n$, $\chi \leftrightarrow m^2 \chi$, $r_0 \leftrightarrow m^2 r_0$ keeping $\varphi$, $n_0$ and the corresponding $n_i$ the same. To overcome the problem of incommensurate length scale it is enough to find a whole number $m'$ such that $m'L \mod k = 0$ and replace the initial arrangement of size $L \times L$ by the periodically created one of size $m'L \times m'L$. Then we can define $S_\Delta(k; L \times L, n, \chi) \equiv S_\Delta(k; m'L \times m'L, m'^2 n, m'^2 \chi)$.

All the above properties can be easily extended to the case of a rectangular image. On the other hand one more interesting feature of the measure is not dependent on the length scale $k$. Namely, the measure value does not change under the replacement "black phase" ↔ "white phase", that is $S_\Delta(k; L \times L, n, \chi) = S_\Delta(k; L \times L, \chi_0 - n, \chi)$. So, for the considered binary image the spatial disorder for a state with a given concentration $\varphi$ of black pixels is the same as for the "inverted" state with the concentration $1 - \varphi$ of white pixels.

Now, for the periodic pattern $mL \times mL$ we shall briefly consider in the thermodynamic limit $m \to \infty$ the reverse situation, i.e. for $k = $ const, $n' \equiv m^2 n \to \infty$ and $\chi' \equiv m^2 \chi \to \infty$ in such a way that $n'/\chi' k^2 \equiv \varphi = $ const. Assuming the large enough length scale $k$ justifying application of Stirling's approximation to formula (1), writing $n_i = (n_i/k^2)k^2 \equiv \varphi_{n_i} k^2$, where $\varphi_{n_i}$ refers to the local particle fraction in $i$th cell, and then using Lagrange multiplier method one obtains

$$S_\Delta(\varphi) \cong \lim_{m \to \infty} \left[ (S'_{max} - S') / \chi' \right]$$

$$= -k^2 \left\{ \varphi \ln \varphi + (1-\varphi)\ln(1-\varphi) - \sum_{n_i} F_{n_i}(\varphi) \left[ \varphi_{n_i} \ln \varphi_{n_i} + (1-\varphi_{n_i})\ln(1-\varphi_{n_i}) \right] \right\}, \quad (5)$$

where the fraction $F_{n_i}(\varphi) = \lim_{m \to \infty} [m^2 g_{n_i}(\varphi)/m^2\chi] = g_{n_i}(\varphi)/\chi$ is related to the frequency of appearance of $n_i$ particles in $i$th cell and $g_{n_i}(\varphi)$ refers to the number of $\varphi_{n_i}$th cells for the initial pattern of size $L \times L$. The work employing this formula in the coarsened lattice model of random granular systems [20] with the known $F_{n_i}(\varphi)$ fractions will be published elsewhere.

## 3. Numerical examples

Here we will consider a few simple patterns to show the entropic measure sensitivity to clustering processes, also to gain insight into the spatial disorder of some fractals and system of interacting particles. As the first example, let us show in Figs. 1a and b two specific patterns of $n = 180$ one-pixel particles grouped into 15 and 5 clusters in linear size $L = 25$ square grid. Each of the bigger clusters is composed of three smaller ones. In Fig. 1c the values of $S_\Delta(k)$ corresponding to the smaller particle clusters show the bottom solid line, while the upper solid line refers to the "clusters of smaller clusters" case. Both the first peaks of $S_\Delta(k)$ are well shaped and confirm the increasing tendency towards the clustering process. As expected, for the bigger clusters the shift in the position of that first maximum, in this case $k = 4 \to 6$, appears. Additionally, the corresponding values of the relative mathematical measure $h_A(k) - h_{Amin}(k)$ [13], dashed lines, are presented (for more details we refer the reader to Appendix B). For this measure slightly different shift in the position of the first maximum appears, i.e. $k = 3 \to 6$. The observed strong similarity between the strictly mathematical and physical entropic measure originates in the formula (2) which is explored in different ways by the two approaches. It should be noted here that the mathematical measure for finite-sized as well as point objects is not invariant under periodical repetition of an initial pattern discussed above in point 6. Nevertheless, its variability along the increasing size of the final patterns is within acceptable range (more details can be found in [19]).

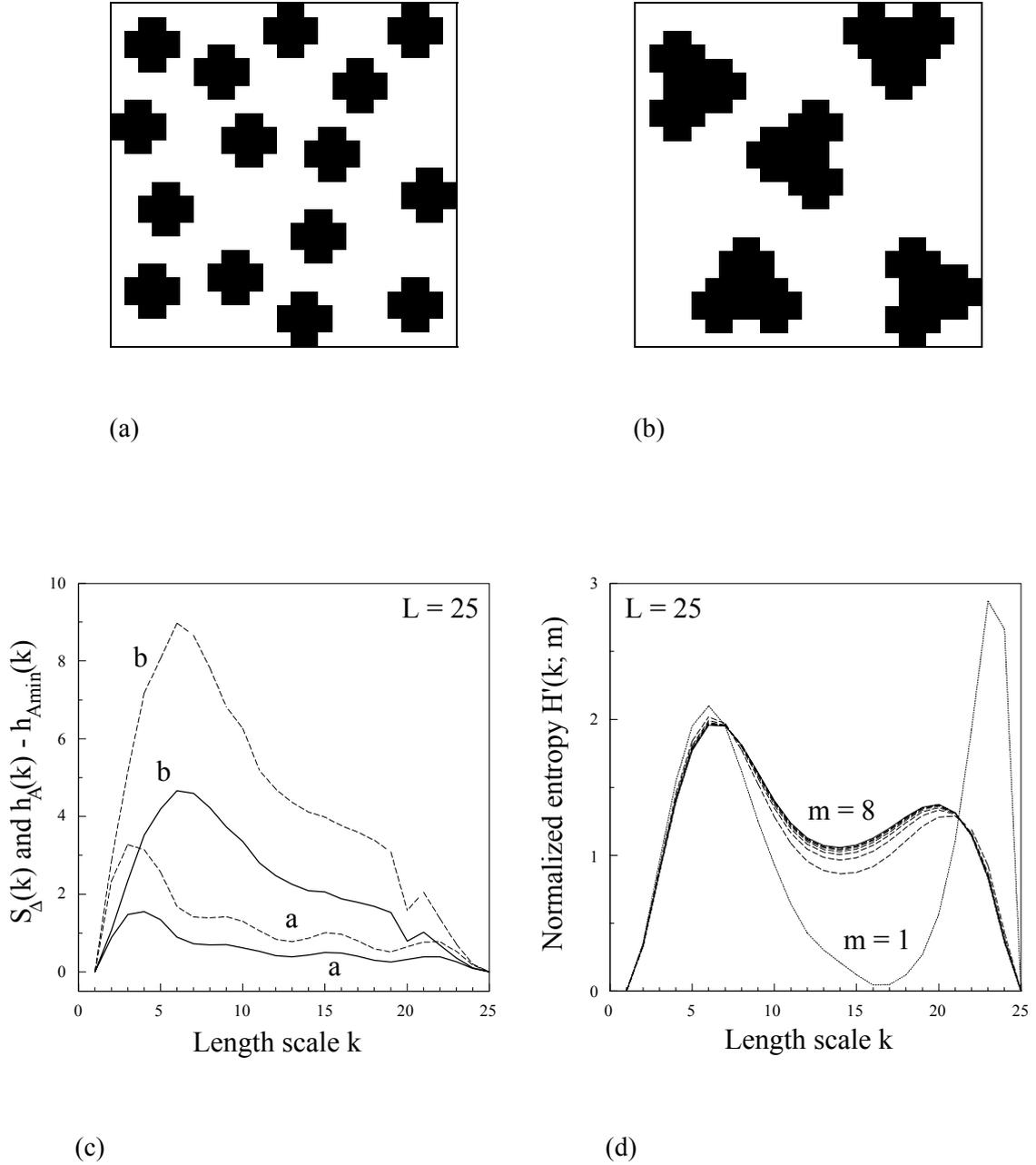

**Fig. 1.** Computer generated cluster distributions consisting of *n* = 180 black pixels and corresponding numerical results for all length scales *k*. (a) The first configuration of 15 identical clusters each composed of 12 pixels. (b) The second configuration of 5 identical clusters each composed of 3 clusters, each of 12 pixels. (c) The entropic measure $S_\Delta(k)$, solid lines, and the relative mathematical measure $h_A(k) - h_{Amin}(k)$, dashed lines, for the two configurations. (d) The sequence of the normalized entropy $H'(k; m)$ curves for $m = 1, ..., 8$ with the representative $H'(k; 8)$ curve, solid line, for the second cluster arrangement. (e) Comparison of the entropic measure $S_\Delta(k)$, solid line, and the $H'(k; 8)$ curve, dashed line, for the first cluster arrangement. Additionally, $H'(k; 1)$ curve, dotted line, is shown. (f) Same for the second cluster arrangement.

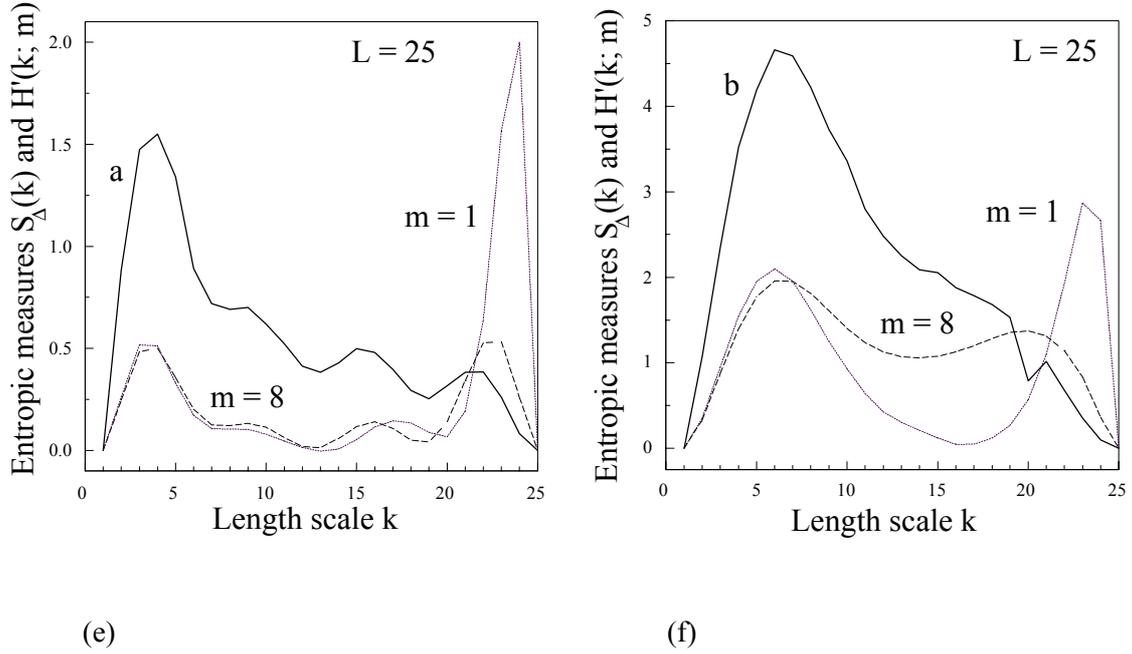

(e)  (f)

Fig. 1. Continued.

In turn, for comparison purposes also the normalized information entropy $H'(k) = H(k) - H_r(k)$ is calculated. However, to estimate the average value of the "experimental" information content, i.e., $H(k)$ the sliding sample box of size $k \times k$ is used. Then, as it was pointed out [18], at scales approaching the system size the deviations of $H'(k)$ are generally largest. To suppress them, Van Siclen's remark [18] about considering the $L \times L$ system to be infinitely periodic rather than finite is employed. Consequently, $H(k)$ values depending on actual arrangement were calculated over $mL \times mL$ periodic system using $k \times k$ sliding box with $k = 1, 2, ..., L$ while the "theoretical" $H_r(k)$ values, as usually, for perfectly random system of size $L \times L$. Such a procedure improves the statistics of sampling as the box size $k$ becomes comparable to $L$. To clarify this matter in Fig. 1d the sequential $H'(k; m)$ curves with the increasing $m$ for the bigger clusters (from Fig. 1b) are presented. The convenient notation $H'(k; m) \equiv H(k; m) - H_r(k)$ is proposed. For the case considered, the representative $H'(k; 8)$ curve, solid line, only slightly differs from $H'(k; 7)$. In Figs. 1e and f the corresponding curves $S_\Delta(k)$, solid lines, $H'(k; 8)$, dashed lines and $H'(k; 1)$, dotted lines, are compared. The first peaks of $H'(k; 8)$ for the two cluster configurations appear at scales $k = 4$ and 6 in agreement with $S_\Delta(k)$. According to definition of $H'(k; m)$ at such scales the particle distribution is more disordered than occurs for a perfectly random configuration while $S_\Delta(k)$ relates the spatial disorder to the appropriate RCM. Despite of their different construction, for these simple cluster configurations the two measures behave at small scales very similarly. For both measures the first peaks can be interpreted as indicating the clustering processes.

Now we focus on self-similar patterns. Among them the Sierpiński carpet family is perhaps the most extensively studied from different viewpoints. For example, on the deterministic Sierpiński carpet (DSC) recent simulation of a phase-separation process has been performed [21] while for the random Sierpiński carpet (RSC) site percolation transition was investigated by a real space renormalization [22], to mention but a few. Using the slightly changed notation of [22] we consider DSC($a$, $b$, $c$) and RSC($a$, $b$, $c$), where an initial square lattice of $L \times L$ cells, with $L = a^c$, is divided into $a^2$ subsquares, only $b$ of them are conserved according to a deterministic rule or at random. The segmentation is repeated on each conserved subsquare, and so on, $c$ times.

Fig. 2a shows DSC(3, 8, 3) and Fig. 2b one of possible RSC(3, 8, 3) both of linear size $L = 27$ and fractal dimension $d_f = 1.89$. Each of these patterns contains $n = 512$ of one-pixel objects. In Fig. 2c the sequence of $S_\Delta(k)$ curves calculated for DSC(3, 8, $c$ = 1, 2 and 3) with corresponding object numbers $n$ = 648, 576 and 512 is presented. This is a fresh picture of the fractals through the filter of our entropic measure $S_\Delta(k)$. It seems that always two peaks at $k = 9$ and 18 dominate while the sequential peaks separated by the interval of size $a = 3$ appear beginning from $k = 3$. For $c = 1$ the relatively strongest disorder appears at scale $k = 18$ while for $c = 2$ and $c = 3$ at scale $k = 9$. Additionally, for the case of $c = 3$, the influence of randomness is presented in Fig. 2d where $S_\Delta(k)$ curves, solid line for DSC and dashed for RSC, show "crossover" behaviour. The crossover range of the length scales depends on realization of random distribution but the values of the peaks in DSC and RSC for $k = 3$, 9 and 18 not. It means that the same occupation numbers $n_i$ appear for the two types of fractals at these scales. This observation agrees with the Sierpiński carpet construction rules. Next, for DSC(3, 8, 3) a comparison of our approach with Van Siclen's results is presented in Fig. 2e. We see that $S_\Delta(k)$ curve, solid line, varies strongly in contrast to smoother fashion of the representative $H'(k; 8)$ curve, dashed line. For illustration of larger deviations also $H'(k; 1)$ curve, dotted line, is shown. The sequential peaks in $S_\Delta(k)$, typical for Sierpiński carpets, are not captured by $H'(k; 8)$. The distinctions appear for the location of the first peak. For $H'(k; m)$ it depends on $m$ in contrast to the cluster configurations previously investigated. For example, now we have $k = 17$ for $m = 1$, $k = 15$ for $m = 2, 3$ and $k = 14$ for $m = 4, ...,8$. In turn, the first peak of $S_\Delta(k)$ measure applied to the Sierpiński carpets always appear at $k = a$, i.e. $k = 3$ in this case. On the other hand, for the scales not being the multiplicity of the parameter $a$, the values of $S_\Delta(k)$ still behave similarly to those of $H'(k; 8)$. It seems indicate that the global shape of $S_\Delta(k)$, that is without sharp local peaks also has some physical meaning. This point needs further investigations.

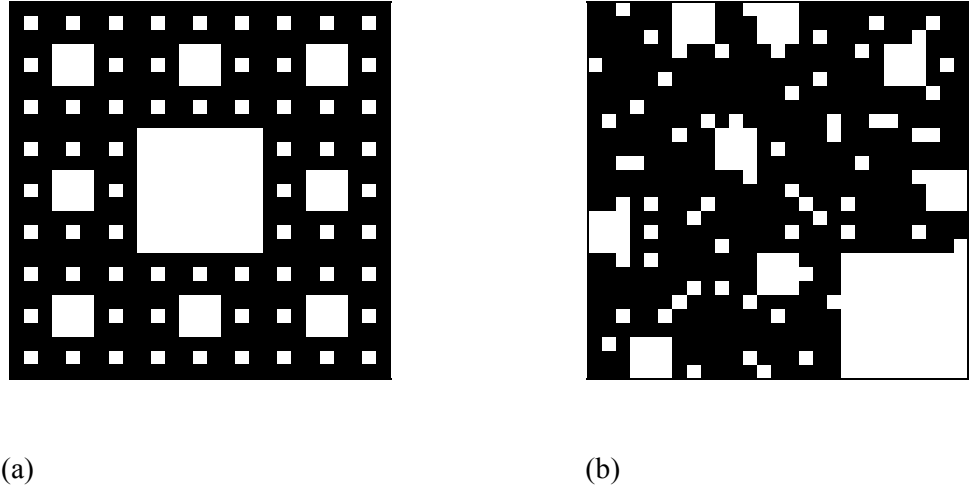

(a) (b)

**Fig. 2.** Examples of the Sierpiński carpets of linear size $L = 27$ and corresponding numerical results for all length scales $k$. (a) DSC(3, 8, 3). (b) RSC(3, 8, 3). (c) The entropic measure $S_\Delta(k)$, solid lines, for DSC(3, 8, $c = 1$, 2 and 3). (d) The entropic measure $S_\Delta(k)$, solid line, for DSC(3, 8, 3) and RSC(3, 8, 3), dashed line. (e) Comparison of the entropic measure $S_\Delta(k)$, solid line, and the $H'(k; 8)$ curve, dashed line, for DSC(3, 8, 3). Additionally, $H'(k; 1)$ curve, dotted line, is shown.

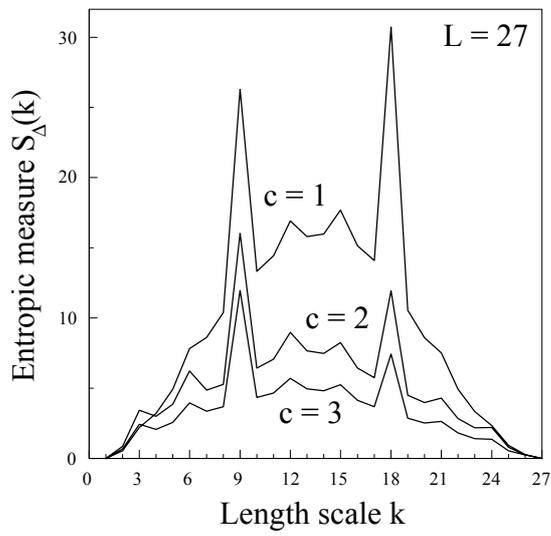

(c)

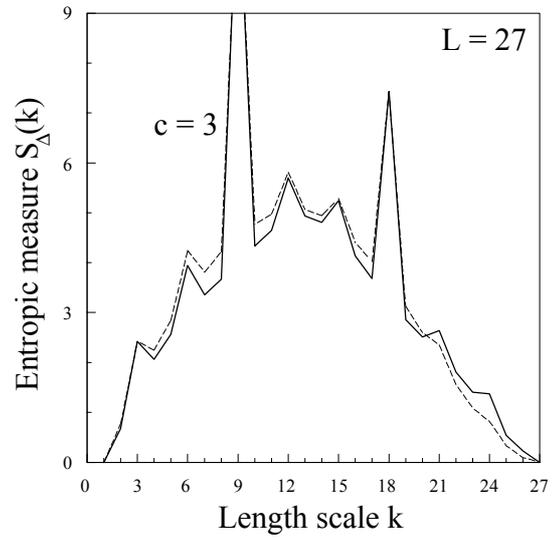

(d)

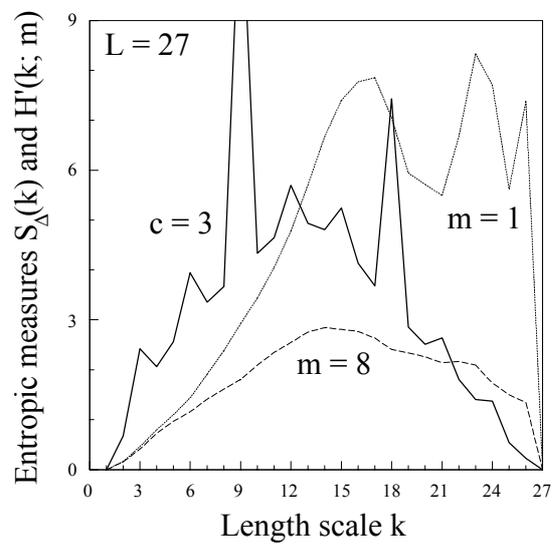

(e)

Fig. 2. Continued.

Let us consider other Sierpiński fractals, for example, DSC(4, 11, 3) and one of possible RSC(4, 11, 3) with fractal dimension $d_f = 1.73$. Each of the corresponding patterns presented in Figs. 3a and b contains $n = 1331$ of one-pixel objects on linear size $L = 64$ square grid. Fig. 3c shows the case that for every length scale the randomness causes the higher spatial disorder measured by $S_\Delta(k)$, dashed line, in comparison with the deterministic fractal, solid line. For both $S_\Delta(k)$ curves the sequential peaks separated by the interval of size $a = 4$ appear beginning from $k = 4$ except the distinguished scale $k = 32$ with a deep local minimum. This is a result of competition between the fractal tendency to clustering at scales being a multiplicity of the parameter $a$ and the prevalent ordering behaviour after division of the pattern into four subsquares with the corresponding actual configuration macrostate {242, 363, 363, 363} and RCM = {332, 333, 333, 333}. The global maximum $S_\Delta(16)$ is relatively very strong. Also for these kind of patterns such specific scales appear, i.e., $k = 4$, 16 and 32, for which the deterministic and random arrangements are not distinguished by $S_\Delta(k)$. In Fig. 3d the comparison of $S_\Delta(k)$, solid line, with the representative $H'(k; 5)$ curve, dashed line, is presented. As previously, also $H'(k; 1)$ curve, dotted line, is shown. We see again that $S_\Delta(k)$ measure varies strongly in contrast to smoother fashion of $H'(k; 5)$ curve. The location of the first peak in $H'(k; m)$ depends on $m$. For example, we have obtained $k = 15$ for $m = 1$, $k = 25$ for $m = 2$, $k = 24$ for $m = 3, 4$, and $k = 23$ for $m = 5$ and 6. Also the similarity between the global shape of $S_\Delta(k)$ and $H'(k; 5)$ is observed.

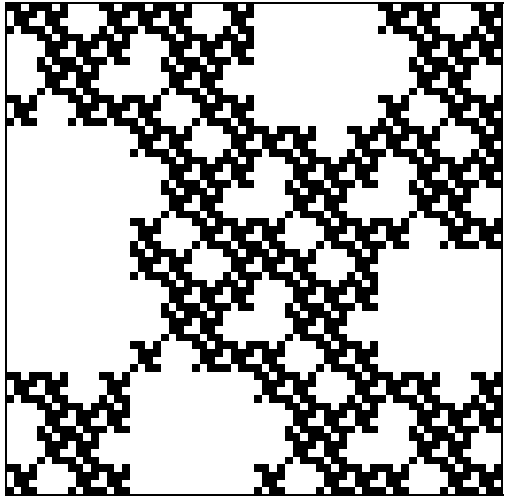
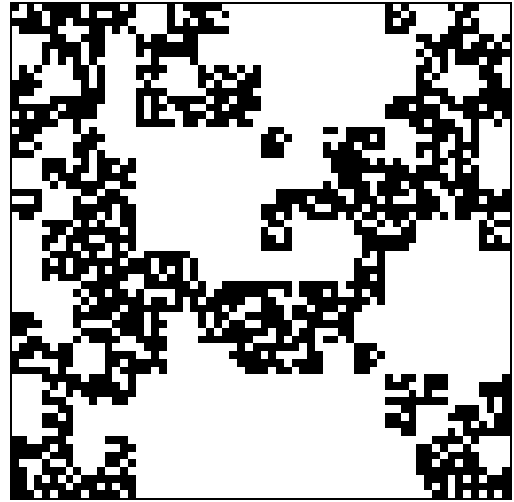

(a)                  (b)

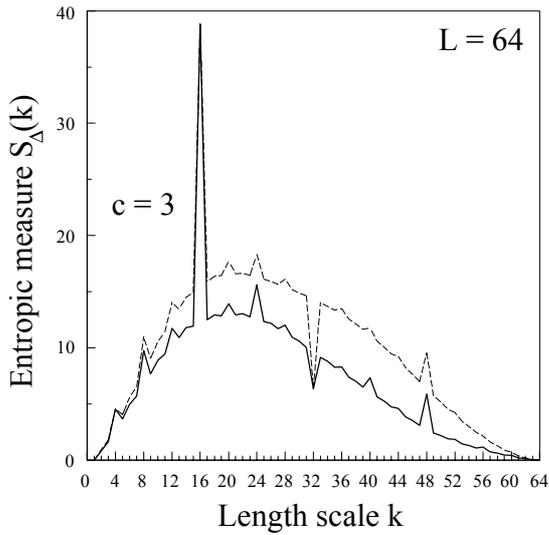
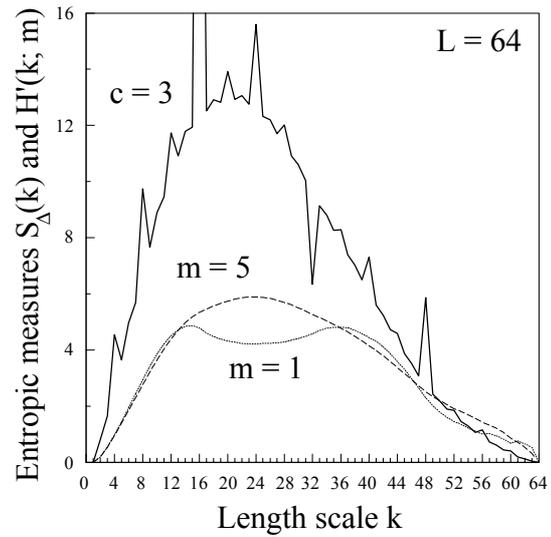

(c)                  (d)

**Fig. 3.** Examples of the Sierpiński carpets of linear size $L = 64$ and corresponding numerical results for all length scales $k$. (a) DSC(4, 11, 3). (b) RSC(4, 11, 3). (c) The entropic measure $S_\Delta(k)$, solid line, for DSC(4, 11, 3) and RSC(4, 11, 3), dashed line. (d) Comparison of the entropic measure $S_\Delta(k)$, solid line, and the $H'(k; 5)$ curve, dashed line, for DSC(4, 11, 3). Additionally, $H'(k; 1)$ curve, dotted line, is shown.

To illustrate the behaviour of the two measures for a population of interacting particles a snapshot from Van Siclen's paper [18] (cf. Fig. 1) has been chosen as the last analyzed pattern. That $n = 625$ configuration of one-pixel objects on linear size $L = 50$ square grid is reconstructed in Fig. 4a. It represents an evolving two-phase microstructure at time $t = 1000$ (more details can be found in Ref. [18]). Fig. 4b shows the corresponding $S_\Delta(k)$, solid line, and $H'(k; 6)$ curve, dashed line. Also $H'(k; 1)$ curve, dotted line, is shown. It can be easily seen that if the first (weakly marked) peak in $S_\Delta(k)$ at $k = 4$ is neglected then the next one (well shaped) appears at $k = 7$ and corresponds to the first peak in $H'(k; 6)$. In contrast to the Sierpiński carpets, now the location of the first peak in $H'(k; m)$ seems to be independent on $m$. The relatively strong global maximum $S_\Delta(25)$ belongs to the sequential peaks of various heights which appear for middle-length scales, approximately for $k = 15$ to $k = 40$. In this case they are separated by the interval of size 5. Such a behaviour is typical to the Sierpiński carpets with a parameter $a = 5$. One may expect that this is a trace of self-similar structure of weakly ramified clusters formed by attracting adjacent particles at the middle-length scales. This possibility has to be further investigated.

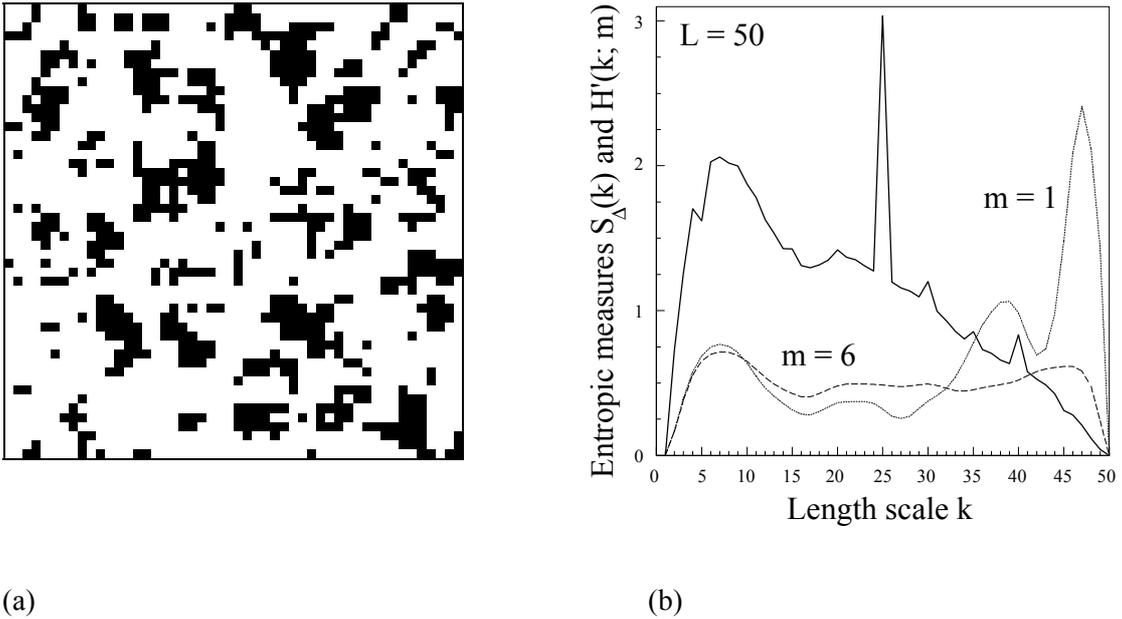

(a)  (b)

**Fig. 4.** Example of a population of interacting particles (adapted from Ref. [18]) and corresponding numerical results for all length scales $k$. (a) Distribution of $n = 625$ interacting particles on linear size $L = 50$ square grid. (b) Comparison of the entropic measure $S_\Delta(k)$, solid line, and the $H'(k; 6)$ curve, dashed line, for the population. Additionally, $H'(k; 1)$ curve, dotted line, is shown.

## 4. Concluding remarks

The present work shows that the relative configurational entropy per cell $S_\Delta(k)$ given by (4) can be considered as alternative, qualitatively correct and highly sensitive measure of spatial disorder at every length scale for systems of finite-sized objects. For the Sierpiński carpets and the population of interacting particle, our measure compared with the normalized information entropy $H'(k)$ reveals additional distinctive features of the complex microstructures. On the basis of results for the fractals, the middle-scale features detected by $S_\Delta(k)$ for the population of interacting particles may indicate for the traces of self-similar structure. For the configurations of compact clusters the two measures are comparable while for specific, highly symmetrical patterns (not common for complex microstructures) the simulations performed but not presented here show that $H'(k)$ provides richer information. Thus, $S_\Delta(k)$ specified with respect to the most spatially ordered configuration and $H'(k)$, which refers to the perfectly random state, can be treated as the two measures providing supplementary information, especially for middle-scale features.

At least two open problems seem to be interesting. First, how make the comparative analysis of binary images of different number of equally sized objects? The example of such a system is by evolving one with the variable number of particles. Concerning the strictly mathematical approach the standardized version of $h_A(k) - h_{Amin}(k)$ is necessary. Second problem is related to quantifying the spatial disorder for the systems of finite-sized objects with a given particle size distribution. From applied point of view such a measure would be useful in granular media. At the thermodynamic limit, such problems may be partially ameliorated by considering $S_\Delta(\varphi)$ given by the formula (5) and using it sequentially for the length scales we are interested in.

The final remark relates to recently reported [23] the novel framework for the study of iteratively constructed fractals with no explicit reference to the space-filling properties of the sets. Deterministic fractal can be defined as a system satisfying a constraint average information required to specify the structure of the set. Such a new approach may shed some light on the morphological changes, for example, for the sequence of DSCs shown in Fig. 2c.

**Acknowledgment**

I thank Clinton DeW. Van Siclen for helpful correspondence.

**Appendix A**

The basic definitions and formulas of Van Siclen's approach are recalled from Ref. [18]. To avoid misunderstanding the only notation change is using $k \times k$ instead of $m \times m$ for the sampling cell. The initially random $N$ particle distribution on a square grid of side length $L$ is represented by the set of *a priori* probabilities

$$p_i(k) = \binom{k^2}{i}\binom{L^2-k^2}{N-i}\binom{L^2}{N}^{-1}, \qquad (A.1)$$

where $p_i(k)$ is the probability of finding exactly $i$ particles in the $k \times k$ square region, $i$ runs from the larger of 0 and $k^2 - (L^2 - N)$ to the lesser of $N$ and $k^2$. The information entropy $H_r(k)$ for the finite, perfectly random system, is then

$$H_r(k) = -\sum_i p_i(k) \log[p_i(k)], \qquad (A.2)$$

while the information entropy $H(k)$ for a given configuration of particles is

$$H(k) = -\sum_i P_i(k) \log[p_i(k)], \qquad (A.3)$$

where $p_i$ is taken from (A.1), and $P_i(k)$ is the actual probability of finding exactly $i$ particles in any $k \times k$ region sampled from the system. The effects of nonrandom particle distribution are measured by the corresponding normalized information entropy $H'(k) = H(k) - H_r(k)$.

**Appendix B**

For easy reference we recall here the final formula for the modified measure $h_A(k)$ [13],

$$h_A(k) = \frac{\mu_A}{E(\mu_A)} \equiv \frac{\chi(\chi k^2 - 1)}{n(\chi - 1)(\chi k^2 - n)}\left(\Delta_A - \frac{n^2}{\chi}\right). \quad (B.1)$$

where $\mu_A \equiv \Sigma^{\chi}_{i=1}(n_i - n/\chi)^2$, $E(\mu_A) \equiv [n(\chi - 1)(\chi k^2 - n)]/[\chi(\chi k^2 - 1)]$ is the expectation value of the random variable $\mu_A$ and $\Delta_A \equiv \Sigma^{\chi}_{i=1} n_i^2$. When $h_A = 0$, that is for $\Delta_A = n^2/\chi \equiv \chi(n/\chi)^2$, the distribution is perfectly homogeneous at a given scale $k$ while for $\Delta_A = n^2$, $h_A = n(\chi k^2 - 1)/(\chi k^2 - n)$ corresponds to the maximum value of its spatial inhomogeneity. However, it should be noted that such configurations are not always possible for the finite-sized objects at a given length scale $k$. Therefore, to evaluate the degree of inhomogeneity referred to its lowest realizable value $h_{A\,min}(k)$ we have suggested [13] to use the relative measure,

$$h_A(k) - h_{A\,min}(k) = \frac{\chi(\chi k^2 - 1)}{n(\chi - 1)(\chi k^2 - n)}\left[\Delta_A - \chi n_0^2 - r_0(2n_0 - 1)\right], \quad (B.2)$$

where $r_0 = n \bmod \chi$, $r_0 \in 0, 1, ..., \chi - 1$ and $n_0 = (n - r_0)/\chi$, $n_0 \in 0, 1, ..., k^2 - 1$.

The similarity between the $S_\Delta(k)$ and $h_A(k) - h_{A\,min}(k)$ is now more obvious (see Fig. 1c). In the thermodynamic limit, using the same notation for the periodic pattern $mL \times mL$ as in Section 2 and Lagrange multiplier method one obtains that $h_{A\,min}(\varphi) = \lim_{m \to \infty} h'_{A\,min} = 0$ and for the above two measures we have the identical final formula

$$h_A(\varphi) = \lim_{m \to \infty} h'_A = \frac{k^2}{\varphi(1 - \varphi)}\left[\sum_{n_i} F_{n_i}(\varphi)\varphi_{n_i}^2 - \varphi^2\right], \quad (B.3)$$

where $F_{n_i}(\varphi)$ and $\varphi_{n_i}$ have the same meaning as previously.